\documentclass[a4paper]{PoS}

\usepackage{amssymb}
\usepackage{fontenc}
\usepackage{times}
\usepackage{mathptmx}
\usepackage{graphicx}
\usepackage{xcolor}
\usepackage{amsfonts}
\usepackage{amsmath}
\usepackage{bm}

\title{Coupling to Multihadron States with Chiral Fermions}

\ShortTitle{Coupling to Multihadron States with Chiral Fermions}

\author{\speaker{Jacob Fallica}$^a$, Keh-Fei Liu$^a$, Jian Liang$^a$, Gen Wang$^a$, Yi-Bo Yang$^b$\\%\thanks{A footnote may follow.}\\
        $^a$Dept. of Physics, University of Kentucky, Lexington, KY 40506\\
        $^b$Institute of Theoretical Physics, Chinese Academy of Sciences, Beijing 100190, China\\
        E-mail: \email{jfa262@uky.edu}}

\abstract{Chiral symmerty is presumed to be a crucial component in the strong interaction and QCD, but its role in spectroscopy, especially for baryons, has not been fully explored.  Compounding this, chiral fermions are uncommon in lattice calculations due to their expensive nature.  We calculate $\eta\pi$, $K\pi$ and $N\pi$ states with $q\bar{q}$ and $qqq$ interpolation fields at $a=0.114\,\mathrm{fm}$ on a $48^3\times 96$ mixed-action lattice at the physical pion mass, with domain-wall sea quarks and overlap valence quarks.  We study the spectral weights of these states as a function of the valence pion mass, which ranges from $m_{\pi}=115-665\,\mathrm{MeV}$, to be compared with the results from non-chiral clover valence quarks on the same domain-wall lattice in order to examine their non-chiral effects, which are expected to decrease with the lattice spacing.}

\FullConference{The 36th Annual International Symposium on Lattice Field Theory - LATTICE2018\\
		22-28 July, 2018\\
		Michigan State University, East Lansing, Michigan, USA.}

\begin{document}

\section{Introduction}

Over the past century, the principles of symmetry have become increasingly important.  The standard model has, among many other symmetries, a spontaneously and explicitly broken chiral symmetry.  Because it is spontaneously broken, Goldstone bosons like the pion emerge; because it is explicitly broken, these Goldstone bosons are not quite massless.  Therefore, that there are important physical consequences of this symmetry is not in dispute.\newline
\indent However, there are restrictive theorems about the incompatibility of lattice regularization, chiral symmetry and ultralocality.  One of the most common choices of lattice discretization for the quarks, clover fermions, contain chiral symmetry-breaking terms besides the usual quark mass term.  Chiral fermions like domain-wall or overlap solve this issue differently, by demanding a modified chiral symmetry relation that goes to the usual continuum as $a$ vanishes.  This modified relation avoids the no-go theorem.  The downside to fermions with improved chiral behavior like overlap or domain-wall is that they are dramatically more expensive than clover fermions.  So, it is important to know when the use of the former must be preferred over the use of the latter.  At larger pion masses and smaller lattice spacings the differences might be negligible, but it is not known.\newline
\indent Here, we examine some lattice results for certain $q\bar{q}$ and $qqq$-type operators coupling to $\eta\pi$, $K\pi$ and $N\pi$ in several symmetry channels where the relevant two-hadron state is expected to be lower in energy than the single hadron ($a_0(980)$, $K_0^*$ or $\kappa$, and $N^*(1535)$, respectively).  Although any operator should always couple to the lowest-lying state in a given symmetry channel, as a matter of practicality the overlap between the operator and the state desired is not always large enough to guarantee that a plateau appears in the time window where statistical errors are reasonable.  We are looking for quantitative and qualitative differences in the correlators and energies for identical operator choices, for chiral and non-chiral fermions, and to examine the effects of chiral dynamics on the two-hadron lowest-lying state.  This work is done on a $48^3\times 96$ lattice, with a spacing of $a=0.114\,\textrm{fm}$.  All our correlators are wall-source to point-sink, in the zero total-momentum channel, which suppresses two-hadron states with nonzero relative momentum.

\section{Mixed-Action}

In this work, we use a mixed action.  This means that the fermions in the valence sector, used for inversions, are different than the fermions in the sea sector, used for configuration generation.  The sea-fermion type is domain-wall, throughout, with a sea pion mass of $139\,\textrm{MeV}$.  For the valence quarks, we use overlap fermions with 12 valence pion masses ranging from $115-665\,\mathrm{MeV}$ as well as clover fermions with 6 valence pion masses ranging from $140-665\,\mathrm{MeV}$.  Because the masses are different in the sea and valence sectors, this action is technically also partially quenched.  Partially quenched actions have a \textit{unitary point}, where the valence and sea masses are identical.  However, because our partially quenched action is mixed between two different types of fermion, this unitary point doesn't exist.  Instead, we expect to find a region where certain effects become more or less important.\newline
\indent Because the valence and sea sectors are different, we can form a larger variety of intermediate states.  Any quark in an intermediate state could be either \textit{valence}\footnote{Connected to the source and sink by quark lines.} or \textit{sea}\footnote{Disconnected from the source and sink.  Or, in perturbation theory, connected by gluon lines only.}, and so quark-line diagrams can become more or less significant as we vary these masses separately.  For example, vacuum bubbles\footnote{By vacuum bubbles we mean the quark loops from the fermion determinant.  If you include gluon lines, as in perturbation theory, these are not vacuum bubbles.} become less important as the valence mass increases.  Although typically cancellations occur between diagrams, by tweaking the fermion type and mass these cancellations are no longer guaranteed.  One should expect to find \textit{ghost} states in certain channels, when a negative term in the correlator dominates, rather than exactly cancels, a positive contribution.  These ghost states are typically indicative of some kind of two-particle behavior, because intermediate states with two or more particles appear from single-hadron interpolating operators thanks to temporally-backwards-moving valence lines or sea bubbles, as illustrated in Figure 1.\newline
\begin{figure}
\begin{center}
\includegraphics[width=0.6\textwidth]{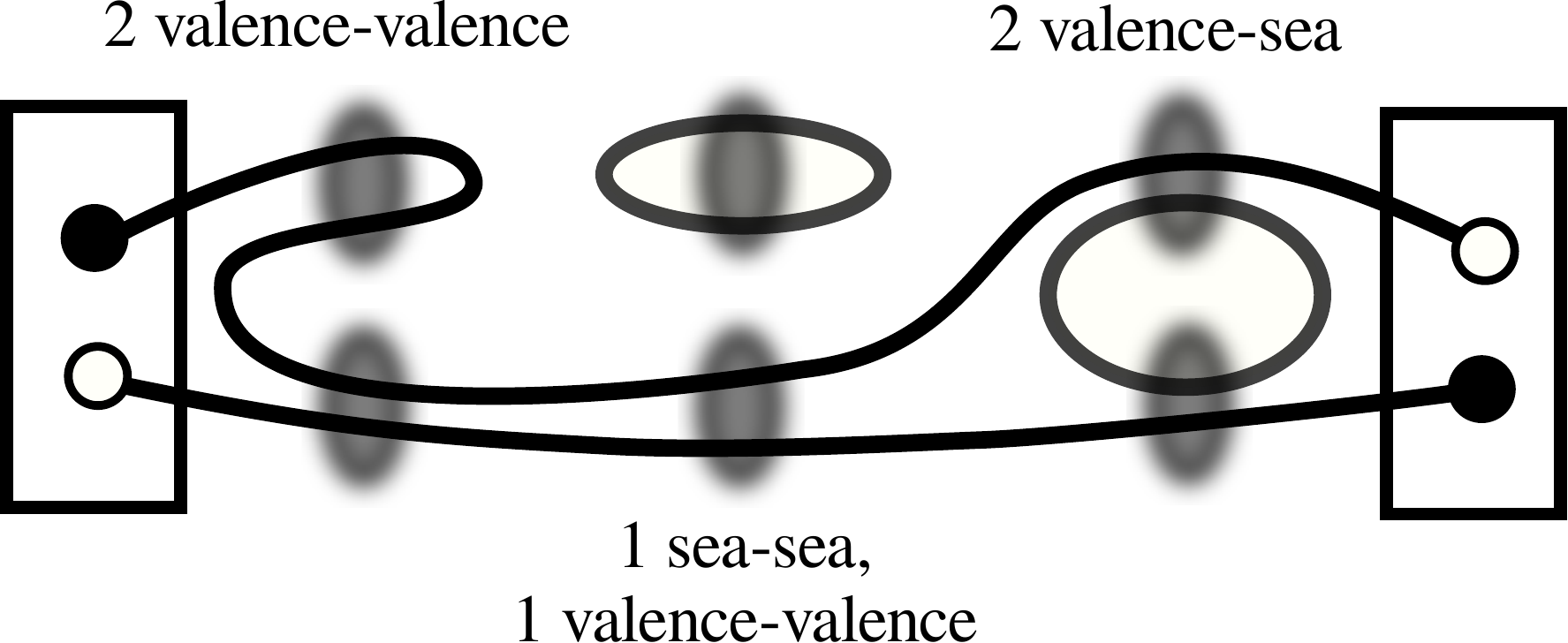}\caption{Diagram of possible contributions to higher Fock spaces in the quark sector.  Illustrated in the diagram from left to right, temporally-backwards-bending lines can give rise to intermediate states made purely of valence quarks, while sea loops can give rise to intermediate states consisting of pure sea and pure valence as well as a mix of valence and sea.}
\end{center}
\end{figure}

\section{Isotriplet Nonstrange Scalar Meson -- $\bm{a_{\bm{0}}}$ Channel}

The lowest state in the $a_0$ channel is the $\eta\pi$ scattering state.  Three classes of path-integral diagrams which are believed to be responsible for producing the $\eta\pi$ state at the unitary poin are illustrated in Figure 2.  For the case of a partially quenched mixed action, the leftmost diagram with sea loops between valence propagators will generate a state of two mixed pions, with mass $2m_{\pi}^{vs}$, where the mixed mass is related to the valence pion mass, sea pion mass and a low-energy constant $\Delta_{\textrm{mix}}$ by the equation\cite{Orginos:2007tw}\begin{equation}
\left(m_{\pi}^{vs}\right)^2 = \frac{ \left(m_{\pi}^{vv}\right)^2 + \left(m_{\pi}^{ss}\right)^2}{2} + a^2\Delta_{\textrm{mix}}.
\end{equation}For the mixed action we are using, $\Delta_{\textrm{mix}}\approx 0.03\,\textrm{GeV}^4$\cite{Lujan:2012wg}.  The middle diagram of Figure 2 is the hairpin diagram which will produce a $\pi\pi$ ghost state by itself in the quenched approximation.  The rightmost diagram, representing a geometric series of loops between the hairpins, when summed with the other two diagrams will produce the $\eta\pi$ state in the unitary case.  For the partially quenched mixed action, we find that when the valence quark mass is smaller than the sea quark mass ($m_{\pi}^{vv}=115\,\textrm{MeV}$ and $130\,\textrm{MeV}$ versus $m_{\pi}^{ss}=139\,\textrm{MeV}$), the middle hairpin diagram dominates and the ghost state drives the $a_0$ correlator below zero, as seen in Figure 3.  For valence quark masses heavier than the sea, the ghost state persists for a bit, until $m_{\pi}^{vv}\approx m_{\pi}^{vs}$, at which point it becomes negligible and soon disappears; Figure 3 shows that overlap valence quarks still see a statistically significant ghost state at $m_{\pi}^{vv}=150\,\textrm{MeV}$.  In contrast, the valence clover fermions do not produce a ghost state at this valence pion mass.\newline
\begin{figure}
\begin{center}
\includegraphics[width=0.8\textwidth]{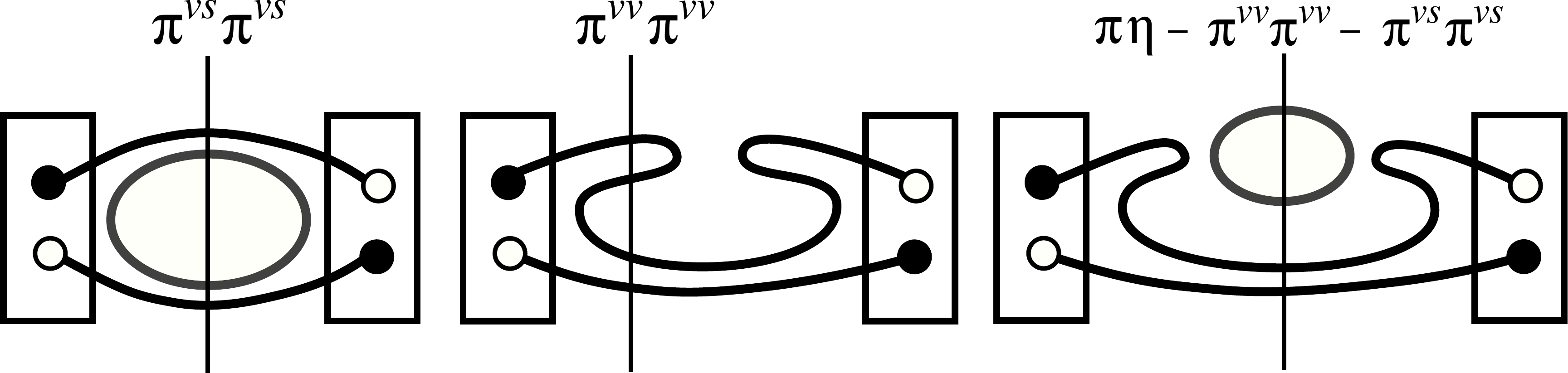}\caption{Diagram of three contributions to a particular correlator.  We can explicitly name the intermediate states, and the additional quarks come from sea loops or from temporally-backwards-bending hairpin diagrams.  Intermediate states are formed as a pair of mixed pions, possibly pure valence pions or a pure valence pion with a pure sea eta.  The rightmost diagram is a part of a geometric series of such diagrams, with infinitely many sea loops; the sum of all these diagrams gives rise to the $\pi\eta$ state.}
\end{center}
\end{figure}
\begin{figure}
\begin{center}
\includegraphics[width=0.7\textwidth]{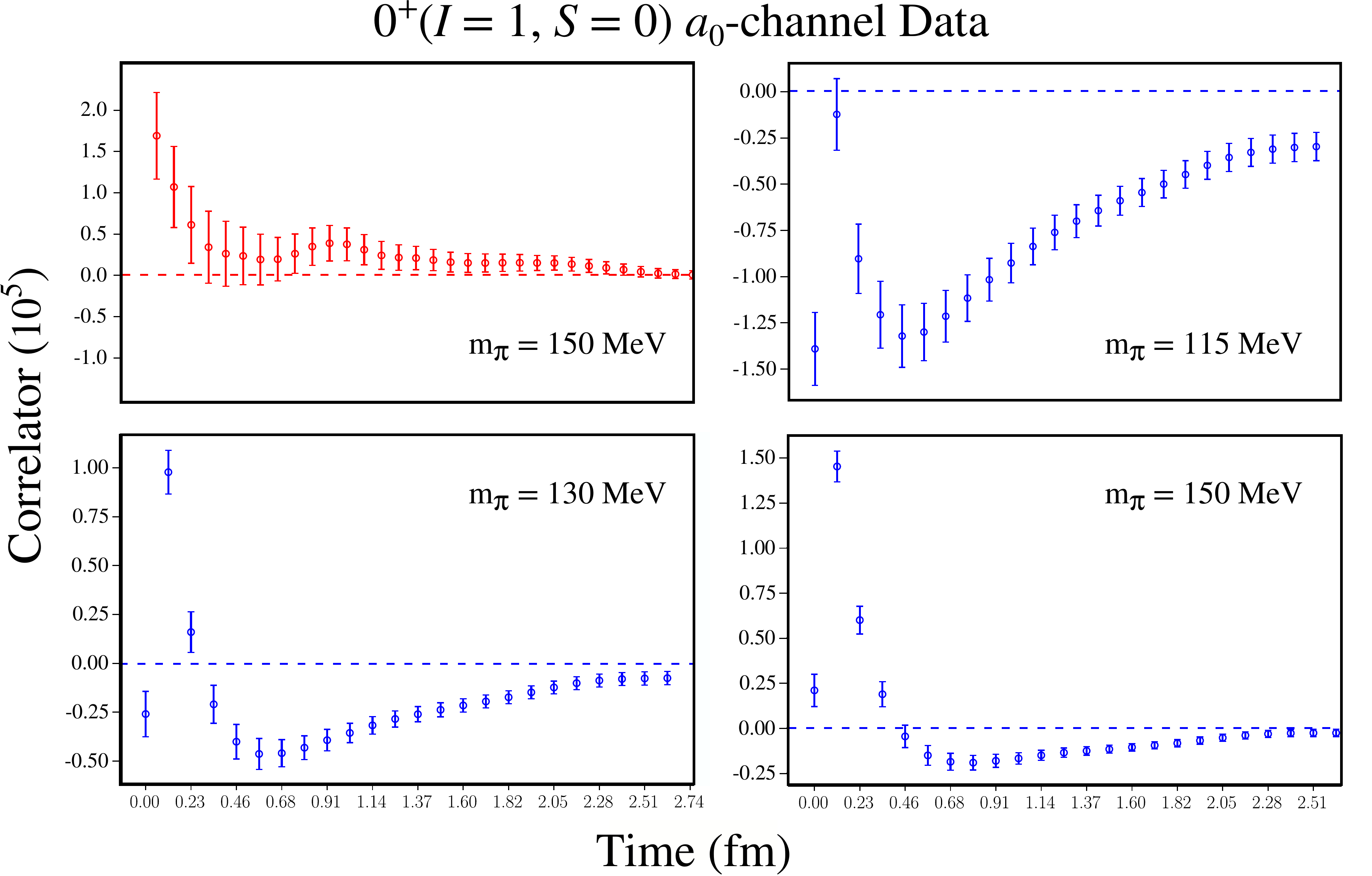}\caption{Four plots of the $a_0-$channel correlator.  The red is clover-on-domain-wall, the blue is overlap-on-domain-wall.  Ghosts are not evident in the top left plot, but are very clear in the other three.  Dotted lines (red or blue) indicate zero.}
\end{center}
\end{figure}

\indent We have calculated the masses in the $a_0$ channel beyond the ghost region, where the physical states appear.  For the region with valence pion mass between $250$ and $400\,\textrm{MeV}$, throughout which $m_{\pi}^{vv}> m_{\pi}^{vs}$, we see from Figure 4 that effective mass of the correlator ranges roughly between $600$ to $700\,\textrm{MeV}$, where we expect to see the $\eta\pi$ scattering state.  Beyond $400\,\textrm{MeV}$, there appears to be a transition, possibly to a state of a different nature; there is a sudden jump in the effective mass and the spectral weight in the upper and lower panels of Figure 4, respectively.  Additionally, results from the valence clover fermions shown on these plots do not demonstrate this feature: above $400\,\textrm{MeV}$ clover and overlap effective masses and spectral weight\footnote{Spectral weights are compared by normalizing according to the central value at our highest valence mass.} are basically the same.  (They diverge for lower valence quark masses, although further studies with both clover valence and sea are needed to confirm this feature.)  It is not clear to us what these states are.  For example, they could also be $\eta\pi$ scattering states, or perhaps they are states dominated by the leftmost diagram in Figure 2, with a repulsive interaction between the two mixed pions.

\begin{figure}
\begin{center}
\includegraphics[width=0.6\textwidth]{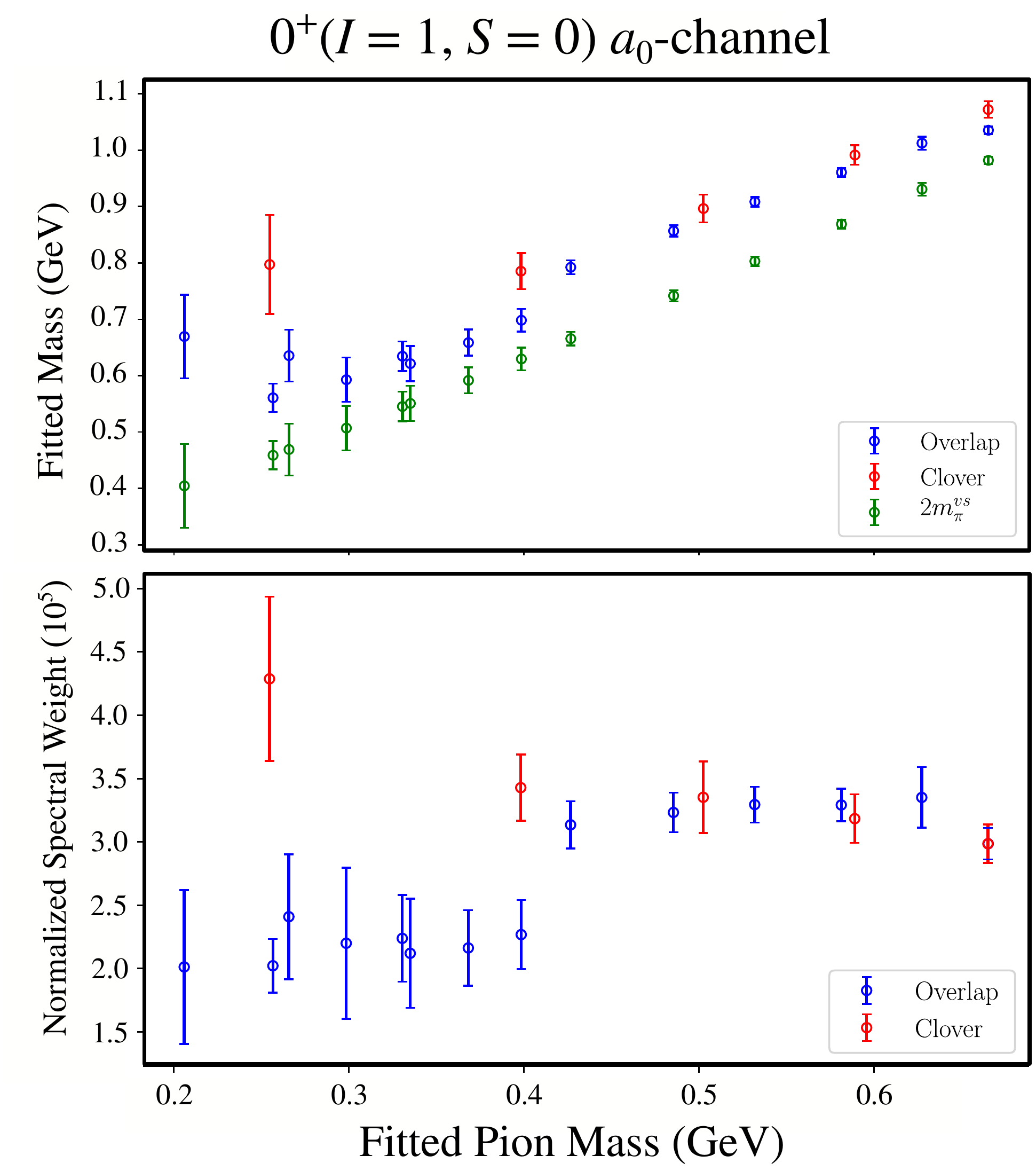}\caption{Energy and spectral weight fit values for the $q\bar{q}$-operator in the $a_0$-channel, as a function of the pion mass.  We compare overlap and clover spectral weight data by normalizing at the highest pion mass, where chiral symmetry is most badly broken.}
\end{center}
\end{figure}

\section{Isodoublet Strange Scalar Meson -- $\bm{K^{\bm{*}}_{\bm{0}}}$ channel}
{\nopagebreak
\begin{enumerate}
\item There is no evidence for the ghost state.  (Figure 8 shows an example correlator.)
\item The results are similar to those in the $a_0$ channel, which contains a lower-mass region below $m_{\pi}^{vv}\approx 400\,\textrm{MeV}$ and, above this, a sudden transition to higher effective masses and overlap factors.
\end{enumerate}}\begin{figure}
\begin{center}
\includegraphics[width=0.6\textwidth]{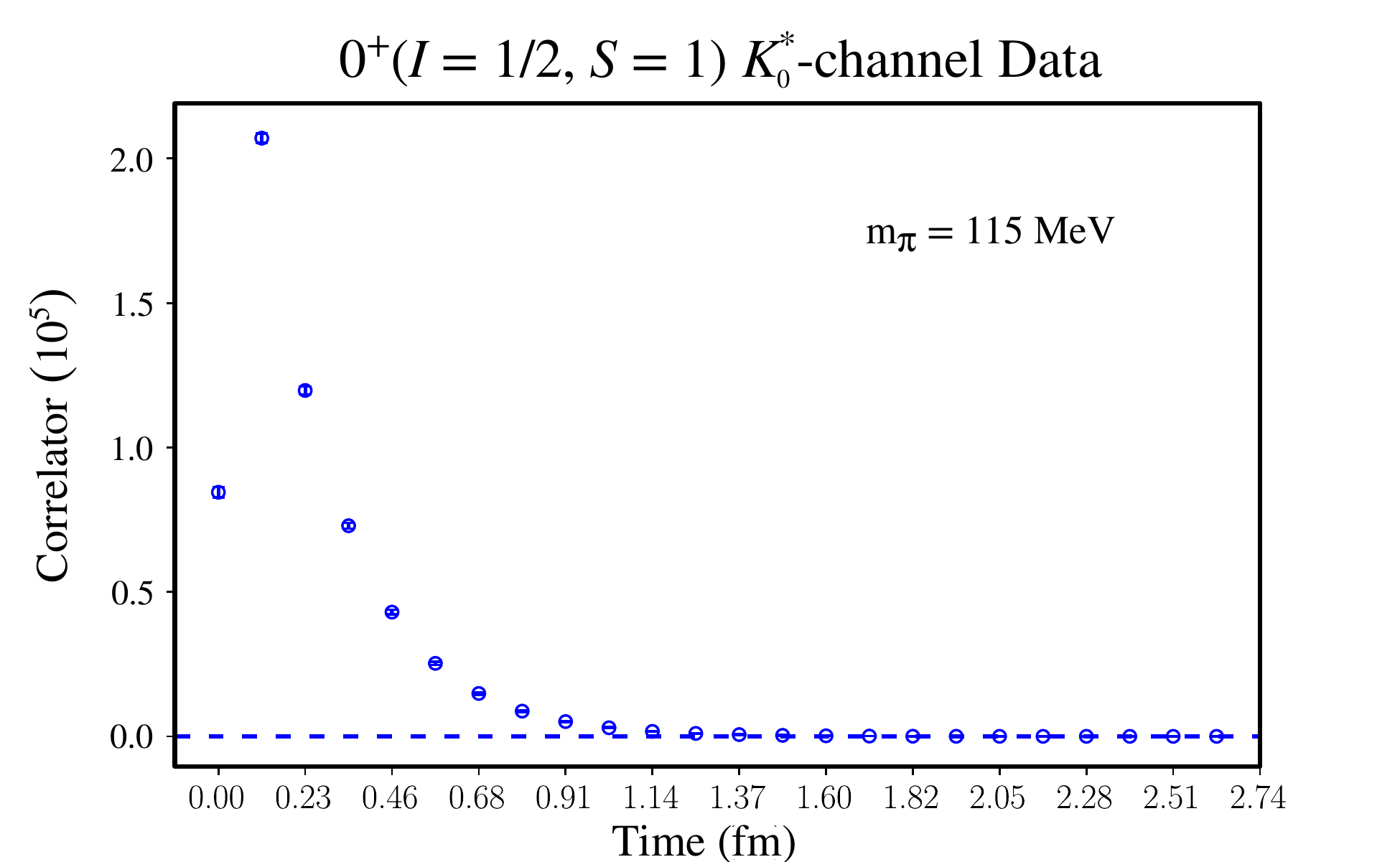}\caption{A plot of the $K^*_0$-channel correlator, for overlap-on-domain-wall.  Ghosts are not evident for the above value $m_{\pi}\approx 115\,\textrm{MeV}$, nor for any others.  Dotted lines indicate zero.}
\end{center}
\end{figure} 
\begin{figure}
\begin{center}
\includegraphics[width=0.6\textwidth]{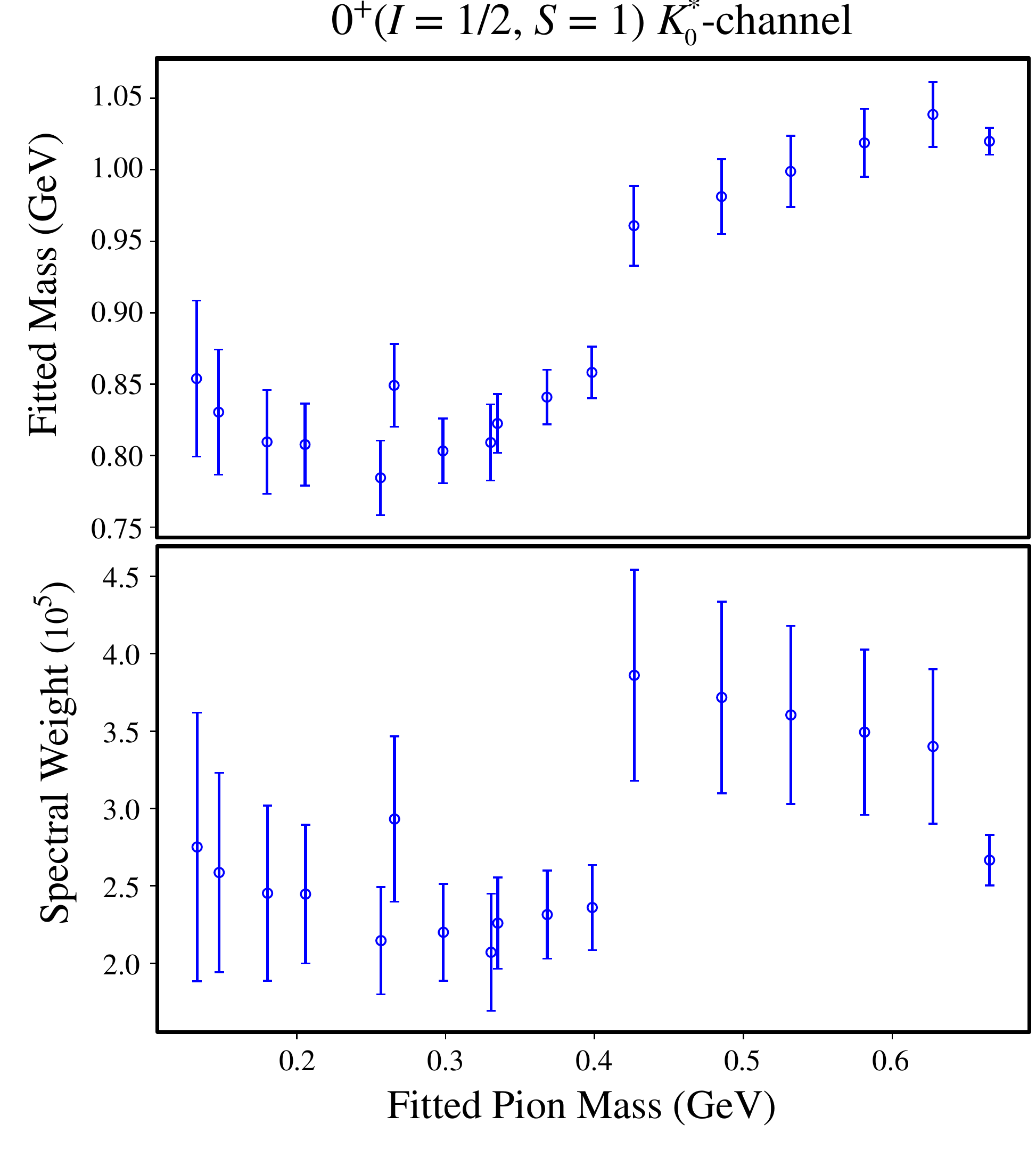}\caption{Energy and overlap factor fit values for the $q\bar{q}$-operator in the $\kappa$-channel, as a function of the pion mass.}
\end{center}
\end{figure}

\section{Isodoublet Nonstrange $\bm{1/2^{\bm{-}}}$ Baryon -- $\bm{N^{\bm{*}}}$ channel}

\begin{enumerate}
\item Three-quark interpolation field is used.
\item Ghost $\pi N$ state is seen for $m_{\pi}^{vv}=180$-$300\,\textrm{MeV}$.  This is shown in Figure 7.  For lower pion masses, the increased statistical error renders any negative central values statistically insignificant, while for higher pion masses the ghost has either disappeared or is indistinguishable from statistical noise at extremely low correlator values (five or six orders of magnitude below the maximum correlator value).
\item In Figure 8, we show the effective mass fits.  Below $m_{\pi}^{vv}\sim 350\,\textrm{MeV}$ the effective mass interpolates between the $\pi N$ state and another, higher state (likely the $N^*(1535)$).  The two states would need to be disentangled with another method, like variational or maximum entropy.
\end{enumerate}
\begin{figure}
\begin{center}
\includegraphics[width=0.7\textwidth]{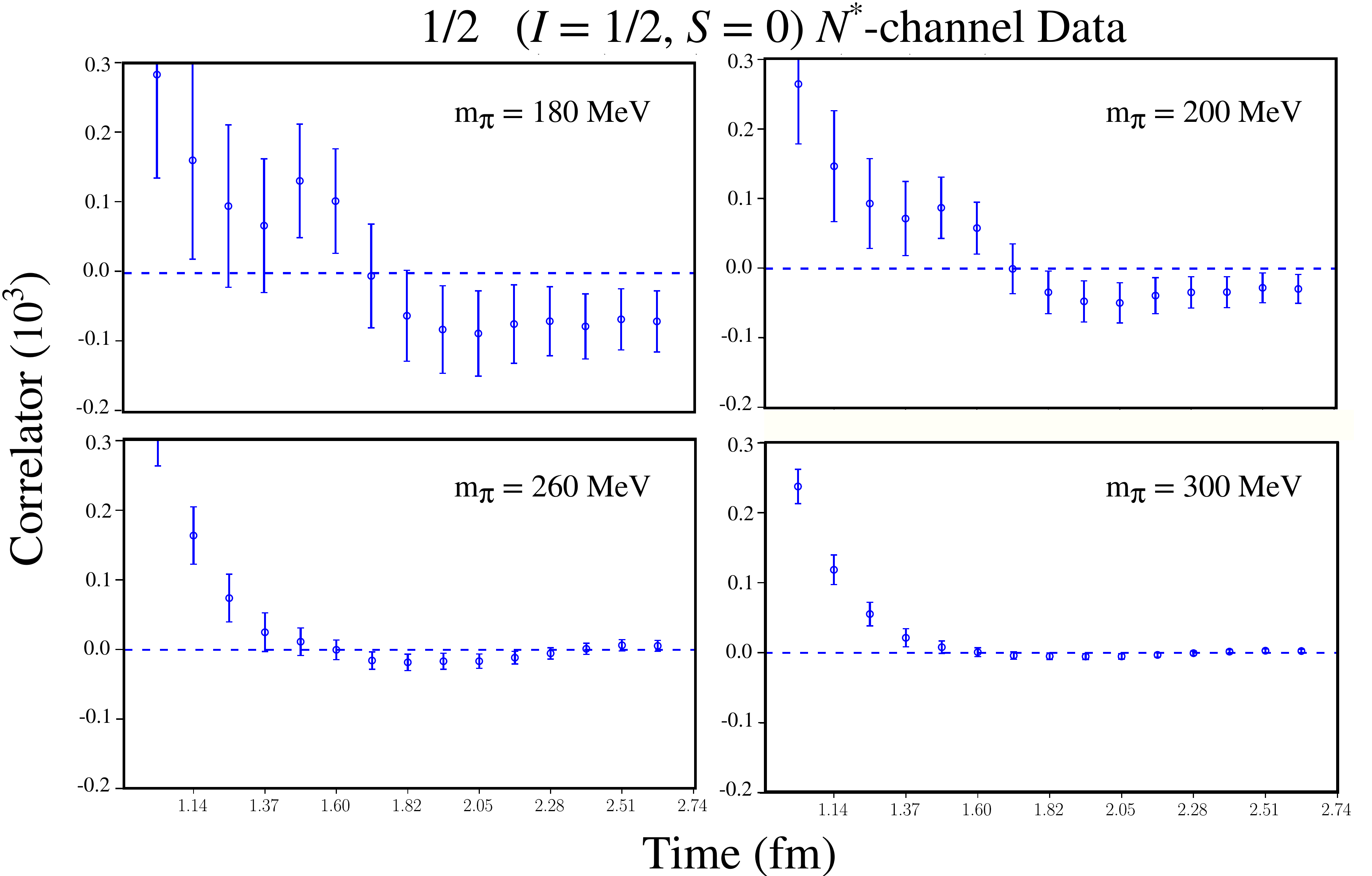}\caption{Four plots of the $N^*$-channel correlator.  All are overlap-on-domain-wall.  Dotted lines indicate zero.  There is a small window from $m_\pi=180$-$300\,\textrm{MeV}$ where ghosts can be reliably seen.  Below $m_\pi=180$, negative central values are rendered statistically insignificant by the larger error bars, while above $m_\pi=300$ the central values are positive or fluctuate around zero with extremely small magnitude.}
\end{center}
\end{figure}
\begin{figure}
\begin{center}
\includegraphics[width=0.6\textwidth]{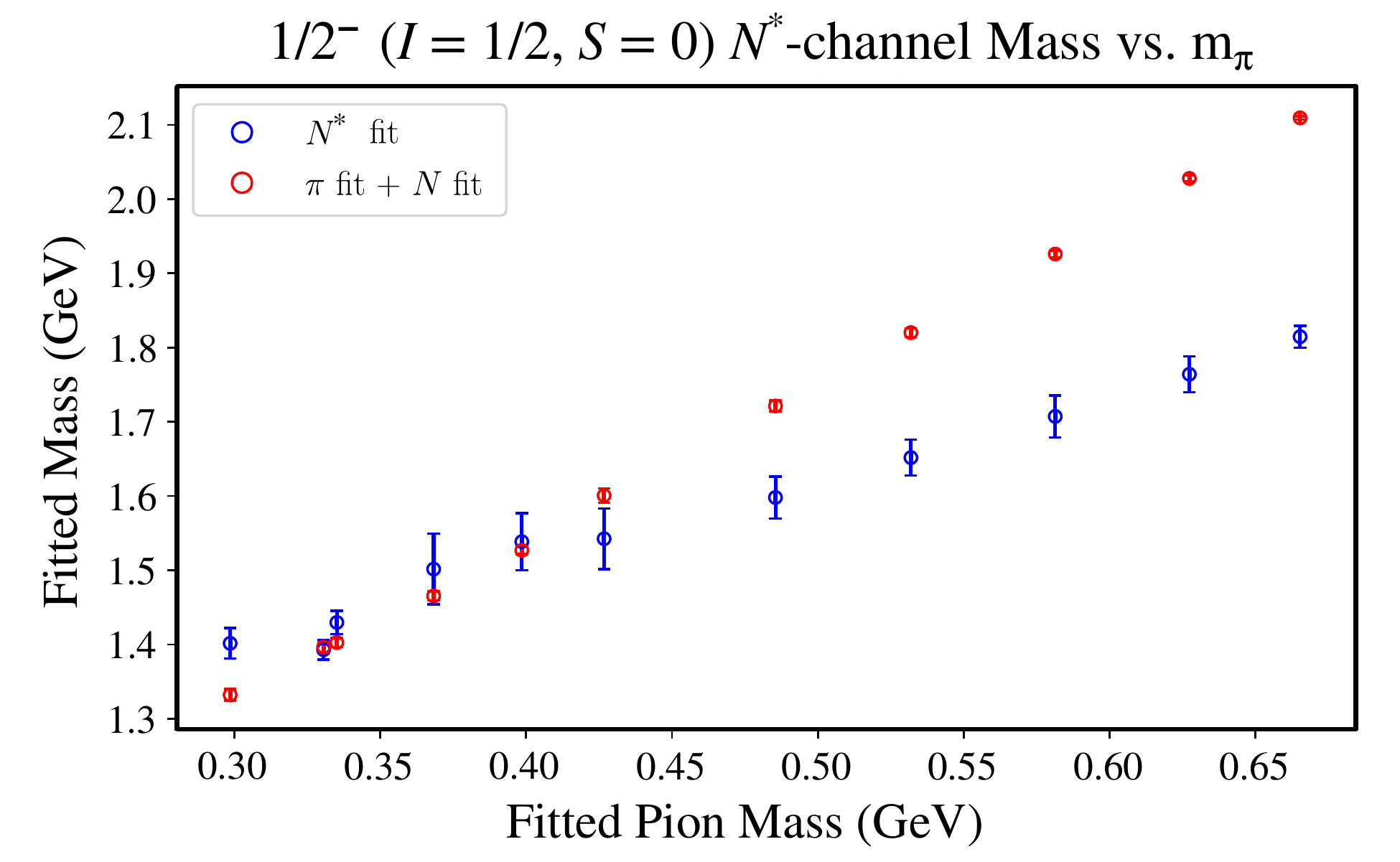}\caption{Energy fit values for the $qqq$-operator, as a function of the pion mass.  We compare the fit values of the $N^*$-channel ground state to the sum of the fit values for the $N$ and $\pi$ states.  Data points below $m_\pi= 300\,\textrm{MeV}$ are omitted, as the significant presence of ghosts renders single exponentials unreliable.}
\end{center}
\end{figure}

\section{Summary}

The multi-hadron states $\eta\pi$, $K\pi$ and $\pi N$, in the $a_0$, $K^*_0$ and $N^*$ channels, are seen for the first time with overlap fermions on domain-wall configurations at the physical pion mass.  Situations like the ghost state in the $a_0$ channel demonstrate differences from the clover case.  This, we believe, reflects some salient features of chiral fermions.  Further studies with a comparison to clover-on-clover results are desirable, in order to disentangle possible mixed-action effects and better understand chiral dynamics in these channels.

\section{Acknowledgments}

This work was supported in part by the U.S. DOE Grant No. DE-SC0013065.  The National Energy Research Scientific Computing Center (NERSC) provided the HPC resources necessary for this research.  Domain-wall fermion gauge configurations were provided by RBC and UKQCD Collaborations.

\end{document}